# Non-expanding universe: a cosmological system of units

Antonio Alfonso-Faus

E.U.I.T. Aeronáutica, Plaza Cardenal Cisneros s/n, 28040 Madrid, SPAIN

E-mail: aalfonsofaus@yahoo.es

**Abstract.** – The product of two empirical constants, the dimensionless fine structure constant ($\alpha$) and the von Klitzing constant ($R_k$, an electrical resistance), turns out to be an exact dimensionless number. Then the accuracy and cosmological time variation (if any) of these two constants are tied together. Also this product defines a natural unit of electrical resistance, the inverse of a quantum of conductance. When the speed of light $c$ is taken away from $\alpha$, as has been shown elsewhere, the constancy of $\alpha$ implies the constancy of the ratio $e^2/h$ (the inverse of the von Klitzing constant), $e$ the charge of the electron and $h$ Planck constant. This forces the charge of the electron $e$ to be constant as long as the action $h$ (an angular momentum) is a true constant too. From the constancy of the Rydberg constant the Compton wavelength, $h/mc$, is then a true constant and consequently there is no expansion at the quantum mechanical level. The momentum $mc$ is also a true constant and then general relativity predicts that the universe is not expanding, as shown elsewhere. The time variation of the speed of light explains the observed Hubble´s red shift. And there is a mass-boom effect. From this a coherent cosmological system of constant units can be defined.

PAC Numbers: 98.80.-k, 98.80Qc

## 1.- Introduction.

The possible time variation of fundamental physical constants, cosmological and at the quantum level too, has been extensively treated in the scientific literature. Here we apply theoretical and experimental evidences to arrive at four constancies: a cosmological quantum of resistance, a cosmological quantum of electrical charge, a cosmological quantum of length and a mass rate. The application of the dimensionless factor ~ $10^{-61}$ to the cosmological charge and length gives the quantum units of electronic charge *e* and length (the Planck´s unit of length, a constant).

We make use of the general relation $G = c^3$ that makes possible to derive the Einstein-Hamilton field equations from the action principle, Alfonso-Faus [1]. We also make use of the relation $\varepsilon c = 1$ found elsewhere, $\varepsilon$ the vacuum permittivity, Alfonso-Faus [2]. Then we arrive at the conclusion that neither the quantum world nor the universe are expanding, and that the speed of light is decreasing with time. Conservation of momentum implies then that masses are increasing with time [1].

## 2. – The cosmological quantum of electrical resistance

The fine structure constant is given by the expression

$$\alpha = \frac{e^2}{4\pi\varepsilon\hbar c}$$

where *e* is the electron charge, $\varepsilon$ the vacuum permittivity, $\hbar$ the Planck´s constant and c the speed of light.

It has been shown, Alfonso-Faus [2], that the relativity principle implies that the fine structure constant $\alpha$ does not contain explicitly the speed of light c. Its expression is found to be

$$\alpha = \frac{e^2}{2h} \qquad (1)$$

This is so because the value of the vacuum permittivity is found to be [2]

$$\varepsilon = \frac{1}{c}$$

The numerical value of $\alpha$ is the same in all system of units, because $\alpha$ is a dimensionless quantity. As of 2007, the best determination of the value of the fine-structure constant is

$$\alpha = 1/137.035999070 \tag{2}$$

Within a time scale of the order of the age of the universe the fine structure constant $\alpha$ appears to be a true universal constant, Chand et al. [3]. In the past the reported observations of a time variation of this constant are meaningless at this scale, Web et al. [4]. The von Klitzing constant $R_k$ [5] is $h/e^2$ so that we get from (1) the result

$$2\alpha R_k = 1 \tag{3}$$

The fact is that the von Klitzing constant, named in honor of Klitzing´s discovery of the Quantum Hall Effect, is the inverse of one quantum of electrical conductance. Its value is given in ohms by the number

$$R_k = \frac{h}{e^2} = 25812.807449 \, \Omega \tag{4}$$

Then from (2), (3) and (4) we obtain a natural unit of electrical resistance as

$$U_r = 2/137.03599907 \times 25812.807449 \, \Omega =$$

$$= 376.73031356 \, \Omega = 1 \tag{5}$$

The von Klitzing constant $R_k$, being the inverse of a **quantum**, strongly suggest by (3) that the fine structure constant can be interpreted as a dimensionless natural quantum too. This supports the idea that the electrical resistance in (5), 376.73031356 $\Omega$, is a natural unit. We also note that this number must be the quantum of electrical resistance of the universe, a cosmological unit as we will see.

### 3. – The cosmological quantum of electrical charge

From the assumed (as an angular momentum) and observed constancy of the Planck´s constant $\hbar$ (radioactive decay times depend on the 7[th] power of $\hbar$, and no time variation in these times has been observed) and (4) one has to assume also that the electronic charge $e$ is a true constant. We have shown elsewhere, Alfonso-Faus [6], that the Planck´s constant for the quantum world is of the order of $1/10^{122}$. We also showed that this constant is the square of the length of the scale, in a system of units that we adopt in part given by the equality $G = c^3$. This equality ensures that the Einstein-Hilbert field equations can be derived from the action principle, Alfonso-Faus [1]. If we take now as the cosmological unit of length the size of the universe ct ≈ $10^{28}$ cm, the cosmological "quantum" of length, then Planck´s constant for the quantum world is the square of the ratio of ct to the Planck´s unit of length

$$L = ct \approx 10^{28} \text{ cm} = 1$$

$$\hbar \equiv (ct)^2 c^3/G\hbar \approx 1/10^{122} \quad (6)$$

and from (1), (2) and (6) we get for the cosmological quantum of electrical charge

$$Q \approx 3.3 \cdot 10^{61} \, e = 1 \quad (7)$$

This is the same value as has been obtained elsewhere, Alfonso-Faus [7], where we have found equilibrium in the plasma state of the universe with charge (7) pushing outwards against the inward gravitational force. The Debye length of this plasma is a bit larger than the size L in (6). We see that the number of electronic charges in (7) is the same as the number of Planck´s units that could make up for the universe (with a dimensionless scale factor of ~ $10^{61}$).

**4.- The cosmological system of units**

We get four cosmological units: length (6), electrical resistance (5), electrical charge (7) and mass rate M/t. The last one comes from the choice of G = $c^3$ that defines a mass rate that is a universal constant. Using the Planck´s units, we have for mass rate:

$$m_p/t_p = (\hbar c/G)^{1/2} \cdot (c^5/G\hbar)^{1/2} = c^3/G \approx 10^{39} \text{gr/sec} = 1 \quad (8)$$

It is interesting to note that if we consider the universe as a black hole of mass M and age t we have

$$GM/c^2 \approx ct \text{ hence } M/t \approx 5 \times 10^{56} \text{gr}/5 \times 10^{17} \text{gr/sec} = 1 \quad (9)$$

This is the same result as in (8). Then we have a cosmological system of four units:

$$\text{Length} \qquad L = ct \approx 10^{28} \text{ cm} = 1$$

$$\text{Electrical resistance} \qquad U_r = 376.73031356 \, \Omega = 1$$

$$\text{Electrical charge} \qquad Q \approx 3.3 \times 10^{61} \, e = 1 \qquad (10)$$

$$\text{Mass rate} \qquad M/t \approx c^3/G \approx 10^{39} \text{gr/sec} = 1$$

We can now define the cosmological Ohm´s law as

$$Mc^2/Q \approx Q/t \times U_r \qquad (11)$$

which is the equivalent electrical potential, energy per unit charge, as the product of the electrical intensity times the electrical resistance. In terms of the Planck´s units and, within a factor of 10, the cosmological Ohm´s law (11) transforms to

$$m_p c^2 t_p \approx \hbar \qquad (12)$$

which is the Planck´s uncertainty principle.

### 5.- A non-expanding universe

From the constancy of the Rydberg constant, $R_y$, Peik et al. [8],

$$R_y = \frac{mc}{4\pi\hbar} \alpha^2 \qquad (13)$$

and the constancy of the fine structure constant α, it is clear that the quantum sizes (proportional to the Compton wavelength $\hbar/mc$) are then constant, no expansion. The momentum $mc$ is proportional to any momentum $mv$, because the ratio $v/c$ must be constant with cosmological time. Otherwise the relativity formulae would be time dependent, which is not observed. Then the constancies of $R_y$, α and $\hbar$, imply the conservation of momentum in the universe. This is Newton´s second law. But under this new view we get a condition from the speed of light: *Any possible cosmological time variation of the speed of light necessarily implies a cosmological time variation of the mass m,* Alfonso-Faus, [1].

According to general relativity the momentum is inversely proportional to the cosmological scale factor R, Harrison [9]. Then the constancy of momentum necessarily implies that the cosmological scale factor is constant. The Universe is not expanding. It is the time variation of the speed of light that causes the observed red shift. There is no theoretical way-out for this conclusion. If one believes in the laboratory measurements and the validity of the general relativity theory *the universe is not expanding and the speed of light is decreasing with time*. Of course the masses are increasing with cosmological time (a Mass-Boom effect, Alfonso-Faus, [1]).

**6. - The Compton wavelength $\lambda_c$.**

It is well known that the Compton wavelength $\lambda_c$ of a mass *m* is given by

$$\lambda_c = \frac{\hbar}{mc} \qquad (14)$$

and that this length gives an order of magnitude of the size of the object of mass *m*. Since theoretical arguments, as well as observations, give constancy for $\hbar$ with cosmological time, from (13) we arrive at the conclusion that the Compton wavelength is also a true constant. The quantum world is not expanding. And the same happens with the de Broglie wavelength $\hbar/mv$.

Harrison [9] has presented a very interesting thought experiment, the so called cosmic-box. The important point here is that from this experiment as well as general relativity, Harrison [9], the momentum of a quantum particle *mv* is inversely proportional to the cosmological scale factor R:

$$mv \sim \frac{1}{R} \qquad (15)$$

There is no way-out of the conclusion that the constancy of momentum forces the constancy of the cosmological scale factor R, and therefore <u>*the universe is not expanding.*</u>

It has been pointed out, Xiaochun [10], that the Robertson-Walker metric leads to the Galileo´s addition rule of velocity. Then it is unsuitable to use this metric to describe universal processes with high speed, e.g. high red shift supernovae. Since the velocity of light has nothing to do with the velocity of the light´s source, the only way to avoid this conundrum is to cancel the derivative of R(t), i.e. the cosmological scale factor has to be constant. Then the Einstein´s cosmological field equations imply that the so called dark energy density (the lambda term), or the pushing force to cancel gravitation, is an expansive electrical one, Alfonso-Faus [7].

Since R is constant, the wavelengths λ of photons are constant too. The observed red shift from distant galaxies is proportional to the measured frequency of the photons $\nu = c/\lambda$. The usual argument runs as follows: given the observed dependence of ν with distance, and therefore with cosmological time, a constant $c$ implies an increasing λ with cosmological time. This corresponds directly to the interpretation that the universe is expanding, that the scale factor R, proportional to λ, is increasing with time. In our case, having shown that λ is constant, the only way-out is to ascribe to the speed of light $c$ a decrease with time. Close to the origin of the universe this speed must have been enormous and this makes the well known horizon paradox and others disappear. A new cosmological model appears to be necessary as a new frame of work based on these findings.

## 6. – Conclusion.

Theory and experiment combine to predict, with high accuracy, the constancies of the fine structure constant $\alpha$, the Planck´s constant $\hbar$, the charge of the electron $e$, the Rydberg constant, the von Klitzing constant, the linear momentum $mv$ and the scale factor of the universe R. The red shift from distant galaxies is here interpreted as a result of the decrease of the speed of light. An increase of mass with cosmological time is also the immediate consequence. The stability of a non-expanding universe may be achieved by the equilibrium between expanding electrical forces and contracting gravitational ones, Alfonso-Faus [7].

A cosmological system of constant units of legth, electrical resistance, electrical charge and mass rate, is then defined and may be applied for cosmology as well as for the quantum world, with an appropriate scale factor. The only possibilities left for fundamental constants to vary with cosmological time are time itself, the speed of light c, the gravitational constant $G = c^3$ and the mass of any object, Alfonso-Faus [1].

## 7. – References.